\documentclass[aps,12pt,superscriptaddress,amsfonts,amssymb,amsmath]{revtex4}

\usepackage{graphicx}
\usepackage{epsfig}
\usepackage{makeidx}
\usepackage{epstopdf}

\begin{document}

\title{
Ultra-light and strong: the massless harmonic oscillator and its singular path integral} 

\medskip

\author{G.\ Modanese \footnote{Email address: giovanni.modanese@unibz.it }}

\affiliation{Free University of Bolzano, Faculty of Science and Technology \\
P.za Universit\`a 5, Bolzano, Italy \medskip}

\date{June 29, 2016}

\linespread{0.9}

\begin{abstract}

\medskip

In classical mechanics, a light particle bound by a strong elastic force just oscillates at high frequency in the region allowed by its initial position and velocity. In quantum mechanics, instead, the ground state of the particle becomes completely de-localized in the limit $m \to 0$. The harmonic oscillator thus ceases to be a useful microscopic physical model in the limit $m \to 0$, but its Feynman path integral has interesting singularities which make it a prototype of other systems exhibiting a ``quantum runaway'' from the classical configurations near the minimum of the action. The probability density of the coherent runaway modes can be obtained as the solution of a Fokker-Planck equation associated to the condition $S=S_{min}$. This technique can be applied also to other systems, notably to a dimensional reduction of the Einstein-Hilbert action.

\end{abstract}


\maketitle

\section{Introduction}

For most dynamical systems, the Feynman path integral admits a continuation to imaginary time, and quantum fluctuations with respect to the classical trajectories are bounded. In certain cases, however, the main contributions to the path integral originate from regions of the configuration space which are far from the classical and semiclassical configurations near the stationary point of the action. Situations of this kind have been previously studied by da Luz et al. \cite{PRA,PHD} for systems with strong constraints or boundary conditions, which lead to a drastic discretization of the configuration space. A remarkable application is to systems, like ``quantum paths'', which are of direct practical interest \cite{q-paths}.

In this paper we will show that a similar ``quantum runaway'' occurs in an apparently harmless system (but one for which the configuration space is truly infinite-dimensional): the massless harmonic oscillator. We shall start with some simple physical remarks and then turn to the path integral.

\section{Elementary treatment: classical, stochastic, quantum}

Let us consider the behavior of an harmonic oscillator in the limit in which the mass $m$ goes to zero while the elastic constant $k$ remains finite. We can think of this limit just as a mathematical exercise, or as a way of describing a situation in which a very light particle is bound by a strong potential. In the classical equation of motion, the limit $m \to 0$ does not produce any qualitative change of the dynamics. The frequency of the oscillations tends to infinity, as can be intuitively predicted considering the lack of inertia. The amplitude of the oscillations, however, remains finite; if the particle starts at rest at some distance from the origin, it remains confined within the same distance, while its velocity at the origin tends to infinity, due to energy conservation. 

Let us check if the classical limit of zero mass can be described by setting $m=0$ in the action of the particle from the beginning, thus starting with the lagrangian $L=-kx^2/2$. The equation of motion obtained by minimizing the action is $x(t)=0$, or $x(t)=F(t)/k$ in the presence of an additional external force $F(t)$. This means that the oscillator just follows the external force, or is at rest at the origin if the external force vanishes. Therefore its behavior is qualitatively different from that obtained taking the limit $m \to 0$. We can conclude that the correct procedure is actually the limit $m \to 0$, because a classical non-relativistic dynamics where $m=0$ from the start is ill-defined (for example, it is impossible to define kinetic energy and momentum of the particle; in fact the canonical momentum associated to $L$ is trivially constrained, etc.).

If we introduce some noise into the classical equations, we still obtain reasonable results in the limit $m \to 0$. In the presence of thermal noise, the equipartition principle suggests that the average displacement of the oscillator is finite, independently from the mass, because $\langle kx^2 \rangle = k_BT$, where $k_B$ is the Boltzmann constant; the average velocity tends to infinity as $m \to 0$, since $\langle mv^2 \rangle = k_BT$. More in detail, one can write a Langevin stochastic equation
\begin{equation}
m x''(t) = -kx(t) + m\Gamma(t)
\label{Lang1}
\end{equation}
where $\Gamma(t)$ is for instance a white noise such that
\begin{equation}
\langle \Gamma(t) \rangle =0, \qquad  \langle \Gamma(t) \Gamma(t') \rangle = 2\gamma \frac{k_BT}{m} \delta(t-t')
\label{Lang2}
\end{equation}
(Here $\gamma$ is an amplitude with dimension time$^{-1}$ \cite{Risken}.)
Then one solves the corresponding Fokker-Planck equation and looks at its limit for $m \to 0$ and $k$ finite. Stochastic mechanics has often been  viewed as a bridge between classical and quantum mechanics \cite{FPvsWF}. The solution $W$ of the Fokker-Planck equation gives the probability distribution of the stochastic variables. In this case it is known that the stationary solution is a Boltzmann distribution in phase space, namely
\begin{equation}
W(x,v) = N \exp \left[-\frac{1}{k_BT} \left( \frac{1}{2} mv^2+\frac{1}{2}kx^2 \right) \right]
\label{Lang3}
\end{equation}
where $N$ is a normalization factor. We see from this distribution that in the limit $m \to 0$, $x$ is limited and $v$ is not, as anticipated through the equipartition law.

When we consider the limit $m \to 0$, $k \ const.$ in quantum mechanics, we immediately realize that the behavior of the oscillator becomes singular. Since the frequency tends to infinity, the energy of the ground state also tends to infinity, and so does the difference between the ground state and the first excited state. The average value of $x^2$ in the ground state also tends to infinity, thus the oscillator becomes completely de-localized; in fact we have
\begin{equation}
\langle 0 \left| \hat{x}^2 \right|0\rangle=\frac{2\hbar}{m\omega}=\frac{2\hbar}{\sqrt{km}}
\label{0q0}
\end{equation} 

The same conclusion can be reached by considering the expressions of the operators ${\hat x}$ and ${\hat p}$ in terms of the adimensional creation/annihilation operators $a^\dagger$ and $a$, namely
\begin{equation}
\hat x = \sqrt {\frac{\hbar }{2}} \sqrt[4]{{\frac{1}{{mk}}}}\left( {{a^\dagger } + a} \right); \qquad \qquad \hat p = i\sqrt {\frac{\hbar }{2}} \sqrt[4]{{mk}}\left( {{a^\dagger } - a} \right)
\end{equation}
This also shows that the roles of the kinetic and potential term in the divergence of the energy are completely analogous, since when $m \to 0$  we have
\begin{equation}
\frac{{{{\hat p}^2}}}{{2m}} \approx \frac{1}{{\sqrt m }}; \qquad \qquad \frac{1}{2}k{\hat x^2} \approx \frac{1}{{\sqrt m }}
\end{equation}

Any correspondence with a classical system is lost, since Glauber's coherent states, usually employed to compute the classical limit of the wave function, cannot be defined when $m \to 0$. We could say that the quantum massless harmonic oscillator is a system which does not admit a classical analogue. On the other hand, an application of the quantum massless oscillator as a physical system at the atomic or sub-atomic level is very unlikely. No real physical force can have a potential proportional to $x^2$ up to an infinite distance. This is usually only an approximation valid near the minimum of the potential, and which makes sense as long as the motion is limited in space; but this does not happen when the mass of the particle tends to zero. (We clearly encounter here also an intrinsic limitation of the non-relativistic theory in the description of massless particles.) 

\section{Path integral }

At the mathematical level, the massless oscillator represents an elementary system which allows to explore some puzzling properties of path integrals. To this end,  let us compute for the oscillator the average of the square of an ``intermediate''  coordinate  
 \begin{equation}
\langle x_{int}^2 \rangle=\frac{
\int {d\left[ x \right]{e^{\frac{i}{\hbar }S\left[ x \right]}}} {x_{int}}^2}
{\int {d\left[ x \right]{e^{\frac{i}{\hbar }S\left[ x \right]}}} } 
\label{eq9} 
\end{equation}
More precisely, consider an intermediate time $t_{int}$ ($t_i < t_{int} < t_f$, where $t_i$ is a given initial time and $t_f$ a final time), call $x_{int}$  the value of the coordinate at the time $t_{int}$  and compute $\langle x_{int}^2 \rangle$, with arbitrary boundary conditions  $q_i$ and $q_f$. A quantity of this kind is also called a ``Feynman-Hibbs transition element''; after discretization, the numerator becomes 
the following ordinary multiple integral over the intermediate coordinates $x_1, \dots ,x_{int}, \dots ,x_{N-1}$
\begin{equation}
\int {d\left[ x \right]{e^{\frac{i}{\hbar }S\left[ x \right]}}} {x_{int}}^2 =
\lim_{\delta \to 0} \int {\prod\limits_{k = 1}^{N - 1} {d{x_k}} x_{int}^2\exp \left\{ {\frac{{ik\delta }}{{2\hbar }}\left[ { - x_i^2 - x_1^2... - x_{int}^2... - x_{N - 1}^2 - x_f^2} \right]} \right\}}
\label{eq10} 
\end{equation}
The exponential is completely factorized and all integrals are simplified in the ratio (\ref{eq9}), except the one in the variable $dx_m$. We obtain
 \begin{equation}
\left\langle {x_{int}^2} \right\rangle  = \frac{{\int {d{x_{int}}x_{int}^2\exp \left\{ {\frac{{ - ik\delta }}{{2\hbar }}x_{int}^2} \right\}} }}{{\int {d{x_{int}}\exp \left\{ -{\frac{{ik\delta }}{{2\hbar }}x_{int}^2} \right\}} }} = \frac{\hbar }{{ik\delta }}
\label{res1}
\end{equation}
In the continuum limit $\delta \to 0$ this is divergent. This result is not an anomaly of the continuum limit of our system, because other quantities have finite averages. For comparison, one can evaluate the average of $\exp(-kx_{int}^2)$, which still gives a Gaussian integral; one easily obtains a finite result, which vanishes in the continuum limit. On the other hand, if we compute the average of a quantity like  $x_{int}^2/\left( {1 + x_{int}^2} \right)$, which tends to a finite value for large  $x_{int}$, we obtain a finite and non-zero result in the continuum limit; to see this, one can exploit the known integral
\begin{equation}
\int {ds\frac{{{s^2}\exp \left( { - \frac{i}{2}a{s^2}} \right)}}{{{s^2} + 1}} = \pi {e^{\frac{{ia}}{2}}}\left\{ { - 1 + {\rm{erf}}\left[ {\left( {\frac{1}{2} + \frac{i}{2}} \right)\sqrt a } \right]} \right\} + \frac{{1 - i}}{{\sqrt a }}\sqrt \pi  } 
\end{equation}
and consider that the ``Error function'' erf($s$) is finite for $a \to 0$, and tends to -1. In our case  $a = const. \cdot \delta $, and $\delta \to 0$  in the continuum limit. The normalization factor at the denominator behaves like $\frac{1}{{\sqrt \delta  }}$, therefore for $\delta \to 0$  the term $\frac{{1 - i}}{{\sqrt a }}\sqrt \pi  $  gives a finite contribution and the rest goes to zero. In conclusion, the quantum averages computed with the path integral tell us consistently that each intermediate coordinate $x_{int}$  tends to grow without limit. Quantum-mechanically, the system runs away to large $x$, far from the classical solution.

\section{Zero-modes and coherent contributions}

The crucial question now is: which configurations in the path integral contribute to this divergent result? Certainly not those near the stationary point of the action, $x(t)=0$. Our conjecture is that there exist a set of configurations for which $S=0$, but $x(t)$ is not zero, and actually not limited. These are not stationary points of the action in the functional space, but in spite of that they give a sizeable, ``coherent'' contribution to the path integral, because for them $S$ is equal to its classical value and the factor $\exp(iS/\hbar)$ is constant. We call these configurations ``zero-modes'' of the action. They satisfy the equation $S=0$, i.e., in this case
 \begin{equation}
x'(t) = \frac{k}{2\mu} x^2(t) + \frac{1}{\mu} f(t)
\label{e1}
\end{equation}
where $f(t)$ is an arbitrary function with zero integral over the time interval we are considering and $\mu$ is an arbitrary non-zero constant, namely the coefficient of a term proportional to $x'(t)$ which one is always free to add to the lagrangian.

\begin{figure}
\begin{center}
  \includegraphics[width=10cm,height=7cm]{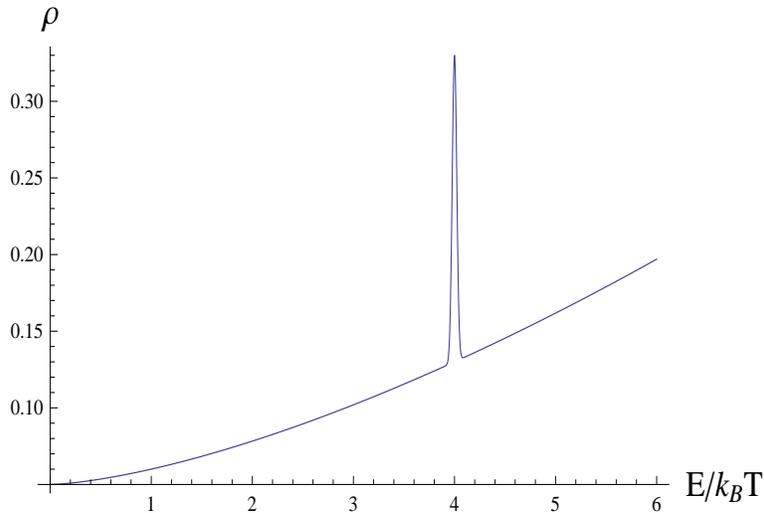}
\caption{Conceptual example of a system with a density of states $\rho(E)$ which has a continuum spectrum plus an integrable spike for some energy value ${\hat E}$. The contribution of the spike to Boltzmann averages is suppressed by the factor $e^{-{\hat E}/k_BT}$. There is no destructive interference in these averages.}
\end{center}      
\end{figure}

\begin{figure}
\begin{center}
  \includegraphics[width=10cm,height=7cm]{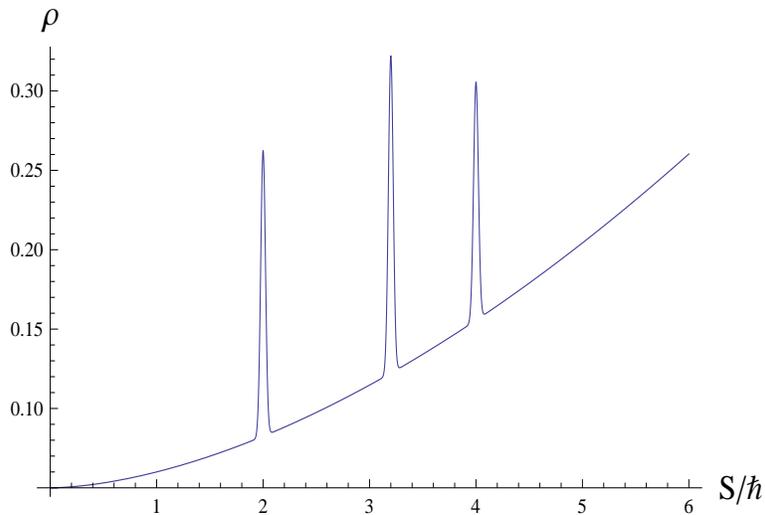}
\caption{Conceptual example of a system with a density of states $\rho(S)$ which has a continuum spectrum plus integrable spikes for some action values ${\hat S}_1$, ${\hat S}_2$, ... The spikes make coherent contributions to the Feynman averages, weighted respectively by factors $e^{i{\hat S}_1/\hbar}$, $e^{i{\hat S}_2/\hbar}$... The contributions from the continuum spectrum interfere destructively between themselves, except for states near the minimum of the action.} 
\end{center}      
\end{figure}

We have proven in \cite{pi} that eq.\ (\ref{e1}) admits runaway solutions which are not limited on any fixed interval and can be properly ``deformed'' through the choice of the function $f(t)$, so that the collection of all these zero-modes of the action makes up a subspace of the configuration space which has non-zero measure; in other words, these configurations make an effective contribution to the path integral and are in fact responsible for the divergence of mean values like $\langle x^2_{int} \rangle$. It can also be proven that, by contrast, the corresponding zero-modes of an oscillator with $m>0$ are limited in any fixed interval.

\begin{figure}
\begin{center}
  \includegraphics[width=10cm,height=7cm]{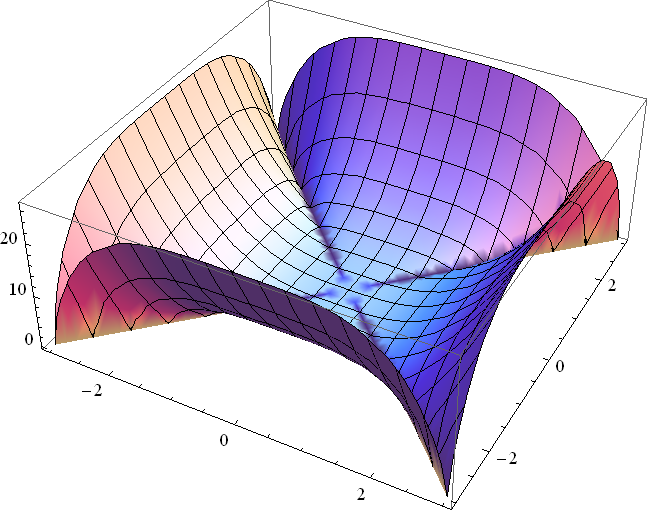}
\caption{Conceptual example of a system whose action depends on two variables $(x,y)$, has a minimum at $(0,0)$ and also stays at its minimum value (but is not differentiable) along two lines crossing at $(0,0)$. In an infinite-dimensional analogue, the ``lines'' of constant action can be a full-dimensional subspace of the configuration space. } 
\end{center}      
\end{figure}

Before giving, in the next Section, a set of stochastic solutions of eq.\ (\ref{e1}), we would like to illustrate more generally the concept of zero-modes and coherent contributions to the path integral. In fact, the application of this concept to the massless harmonic oscillator is only one example, but there are other possible applications (notably to General Relativity \cite{BUC}).

Let us start with a simpler analogy, namely with a functional integral in statistical mechanics with a real Boltzmann weight $e^{-E/k_BT}$. If we know the density of states $\rho(E)$ as a function of the energy, we can write the average value of any function $F(E)$ as
 \begin{equation}
\langle F(E) \rangle = N \int dE \rho(E) F(E) e^{-E/k_BT}
\label{if1}
 \end{equation}
where $N$ is a normalization factor.

If the density $\rho(E)$ has a $\delta$-like peak for some value ${\hat E}$ (Fig.\ 1), then we expect that the configurations with energy ${\hat E}$ give a relevant contribution to the averages. If, however, ${\hat E}$ is large with respect to the ground state energy and the temperature $T$ is not very high, the contribution of the states with energy ${\hat E}$ will be quite suppressed by the factor $e^{-{\hat E}/k_BT}$; and in any case, all values of $E$ contribute to the averages, albeit with the weight $e^{-E/k_BT}$.

In a Feynman path integral with complex weight $e^{iS/\hbar}$ we can imagine to integrate over a density of states $\rho(S)$ which is a function of the action of the states. The states can be thought of as classical trajectories. We can write the average of a function $F(S)$, similarly to (\ref{if1}), as
 \begin{equation}
\langle F(S) \rangle = N \int dS \rho(S) F(S) e^{iS/\hbar}
 \end{equation}
In this case, however, the interference factor $e^{iS/\hbar}$ leads to a mutual cancellation of all the contributions from the continuous spectrum, except for those near the minimum. The other surviving contributions are those from possible $\delta$-like peaks (Fig.\ 2), which can also interfere between themselves, depending on the phase differences between ${\hat S}_1/\hbar$, ${\hat S}_2/\hbar$ ...

In our current example we are considering a system whose density $\rho(S)$ has a minimum {\em and} a peak at $S=0$ (and can also have other peaks, not computed here). A graphical representation which illustrates the behavior of the action $S$ as a functional of the classical paths is given in Fig.\ 3. It is impossible, however, to fully represent in a finite-dimensional image a functional over an infinite-dimensional space. The idea is that the zero-modes are elements of those subspaces where $S$ is zero but not a minimum, like the four lines in the horizontal plane of Fig.\ 3, where the function of two variables represented is not differentiable. The smooth minimum of $S$ is at the origin and is surrounded by a full-dimensional region where $S$ is very close to zero; but in the infinite-dimensional analogy also the ``lines'' of the zero-modes are infinite-dimensional like the full space and do not have zero measure like lines in 2D.

\section{Stochastic and Fokker-Planck equation for the zero-modes}

We would like to offer here an alternative characterization of the zero-modes, based on a correspondence between eq.\ (\ref{e1}) and the over-damped Kramers stochastic equation, namely
 \begin{equation}
y'(t) = \frac{1}{M\gamma} F(y)+ \frac{1}{\gamma} \Gamma(t)
\label{e2}
\end{equation}
where $y(t)$ represents a stochastic process, the parameter $\gamma$ and the noise $\Gamma(t)$ are defined as in eq.\ (\ref{Lang2}) above, and $M$ represents a mass (which in the following does not tend to zero, but is fixed). With the identifications $M=1/k$, $\gamma=|\mu|$, $F(y)=\pm y^2/2$, we see that (\ref{e2}) corresponds to (\ref{e1}), provided we also admit that $\int \Gamma(t)dt=0$ over any interval under consideration. This latter hypothesis (ensemble average equals time average) is very natural for a white noise stochastic process with zero correlation time like $\Gamma(t)$. It follows that the ensemble of stochastic functions $y(t)$ is distributed according to the solution $W(y,t)$ of the following Fokker-Planck equation (Smoluchowski equation)
 \begin{equation}
\frac{\partial W}{\partial t}= \frac{1}{M\gamma} \left[ -\frac{\partial}{\partial y} F(y) + k_BT \frac{\partial^2}{\partial y^2} \right] W
\label{e2-1}
\end{equation}

\begin{figure}
\begin{center}
  \includegraphics[width=10cm,height=7cm]{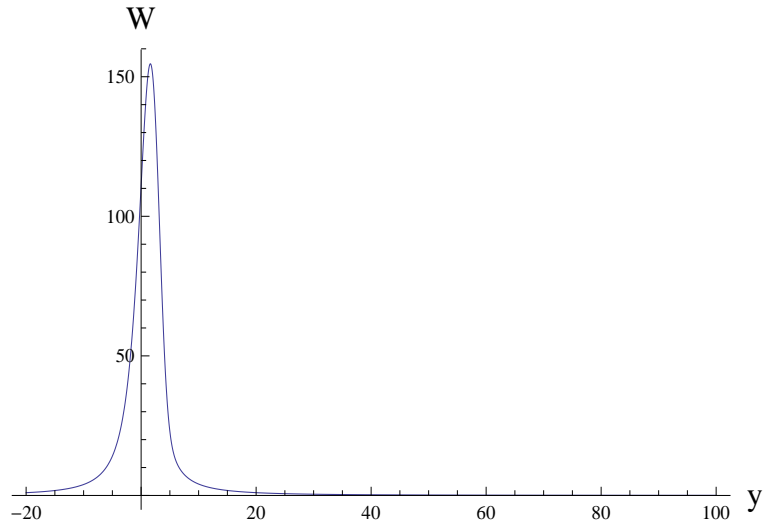}
\caption{A numerical solution of the Fokker-Planck equation (\ref{e3}) with the + sign. In this case the solutions are limited, meaning that the coordinate $x$ of the oscillator has zero probability to escape to infinity. } 
\end{center}      
\end{figure}

\begin{figure}
\begin{center}
  \includegraphics[width=10cm,height=7cm]{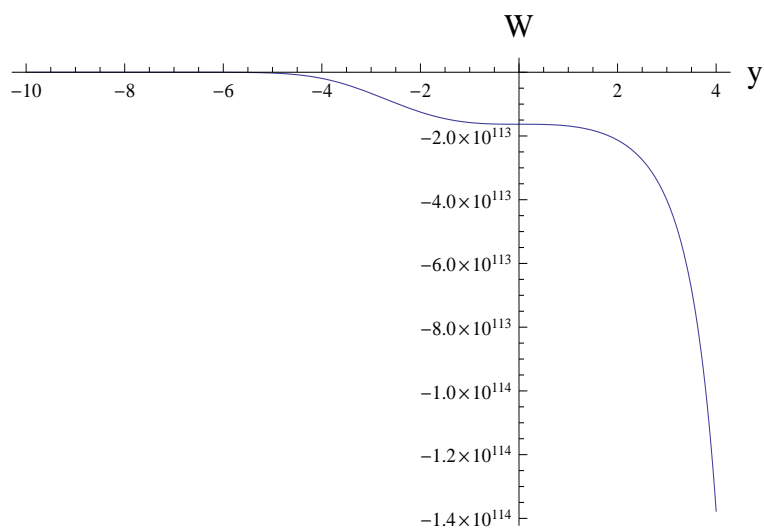}
\caption{A numerical solution of the Fokker-Planck equation (\ref{e3}) with the -- sign. In this case there are non-limited solutions on any interval. We interpret this as due to the quantum runaway which causes the average $\langle x^2_{int} \rangle$ to diverge.} 
\end{center}      
\end{figure}

Let us look for stationary solutions of this equation. We obtain the homogeneous linear ordinary differential equations
 \begin{equation}
\pm y W(y) \pm \frac{1}{2} y^2 W'(y) - k_B T W''(y)=0
\label{e3}
\end{equation}
according to the choice $F(y)=\pm y^2/2$ (note that this choice depends in turn on choosing $\mu$ positive or negative, while the sign of $\Gamma(t)$ is irrelevant). It is straightforward to check that with the + sign eq.\ (\ref{e3}) admits only limited solutions resembling Gaussians $W \sim \exp[-c(y-y_0)^2]$, while with the -- sign the solutions are not limited (see Fig.s 4, 5 for some numerical examples). This confirms the existence of zero-modes of the action which are not limited in space.

\section{Conclusions}

We started our analysis looking at the elementary properties of a massless oscillator, regarded as a classical, stochastic and quantum system. Since the quantum dynamics is readily seen to be singular, we turned to consider the Feynman path integral, in order to clarify from the mathematical point of view the origin of these divergences. We found that the functional averages of quantities which are limited as functions of the coordinate $x$ are finite, therefore the divergence signals a true quantum runaway from the classical trajectories which minimize the action. This points to a more general criterion: if we suppose to perform the functional integration over a ``density of states'' $\rho(S)$ which depends on the action of the configurations, then for systems with a continuum spectrum the interference factor $e^{iS/\hbar}$ ensures that the only relevant contribution originates from the stationary points of $S$; if, however, there are $\delta$-like peaks in $\rho(S)$, then the system can in fact run away to non-classical configurations. We have computed explicitly some of these configurations for the massless oscillator by solving the equation $S=0$, regarded either as a differential equation depending on a function with null integral, or as a stochastic equation containing a white noise. A possible application of this method to a system far more complex and physically relevant than the massless oscillator concerns the Einstein gravitational action $S=(1/8\pi G) \int d^4x \sqrt{g(x)} R(x)$. Since the scalar curvature $R$ is not positive-definite, it is possible to find field configurations which do not satisfy the Einstein equations in vacuum and still have zero action after the integration, due to a compensation between disjoint regions with positive and negative curvature. Configurations of this kind have been computed in \cite{BUC,dip1,dip2}. (See also a related equipotential lines technique in \cite{compl}.) In order to evaluate their contribution to the Feynman path integral one can follow an approach similar to that described here.

\end{document}